\documentclass[cup6b,epsf]{cupbook}

\usepackage{epsf}

\def\note #1]{{\bf #1]}}

\def\note #1]{{\bf #1]}}

\begin{document}

\author{MARIO HAMUY\\
The Observatories of the Carnegie Institution of Washington \\
813 Santa Barbara St., Pasadena, CA 91101, USA}

\chapter{Observed and physical properties of type II plateau supernovae}

{\it I use photometry and spectroscopy data for 24 Type II plateau supernovae to
examine their observed and physical properties. This dataset shows that
these objects encompass a wide range in their observed properties (plateau luminosities,
tail luminosities, and expansion velocities) and their physical parameters
(explosion energies, ejected masses, initial radii, and $^{56}$Ni yields).
Several regularities emerge within this diversity, which reveal
(1) a continuum in the properties of Type II plateau supernovae,
(2) a one parameter family (at least to first order),
(3) evidence that stellar mass plays a central role in 
the physics of core collapse and the fate of massive stars.
}

\section{Introduction}

Type II supernovae (SNe~II, hereafter) are exploding stars characterized by
strong hydrogen spectral lines and their proximity to star forming
regions, presumably resulting from the gravitational collapse of
the cores of massive stars ($M_{ZAMS}$$>$8 $M_\odot$).
SNe~II display great variations in their spectra and lightcurves
depending on the properties of their progenitors at the time
of core collapse and the density of the medium in which they explode.
Nearly 50\% of all SNe~II belong to the plateau subclass (SNe~IIP) which
constitutes a well-defined family distinguished by
1) a characteristic ``plateau'' lightcurve (Barbon et al. 1979),
2) Balmer lines exhibiting broad P-Cygni profiles, and
3) low radio emission (Weiler et al. 2002). These SNe are thought
to have red supergiant progenitors that do not experience significant mass loss
and are able to retain most of their H-rich envelopes before explosion.
In section 1.2 I summarize the observed properties of SNe~IIP based on
a sample of 24 objects, and in section 1.3 I use published models to
derived physical parameters for a subset of 13 SNe. 

\section{Observed properties of Type II plateau supernovae}

In Hamuy (2003; H03 hereafter) I compiled photometric and spectroscopic data
from my own work and a variety of publications, for a sample of 24 SNe~II.
In Table 2 of that paper I summarized observed parameters, such as
the absolute $V$ magnitude near the middle of the plateau ($M^V_{50}$),
the duration of the plateau, the velocity of the expanding envelope
measured near the middle of the plateau ($v_{50}$), and the luminosity
of the exponential tail (converted into $^{56}$Ni mass ejected
in the explosion). The wide range in luminosities and expansion velocities
is clear manifestation of the great diversity of SNe~IIP.

\begin{figure}
\begin{center}
\leavevmode\epsfxsize=10cm \epsfbox{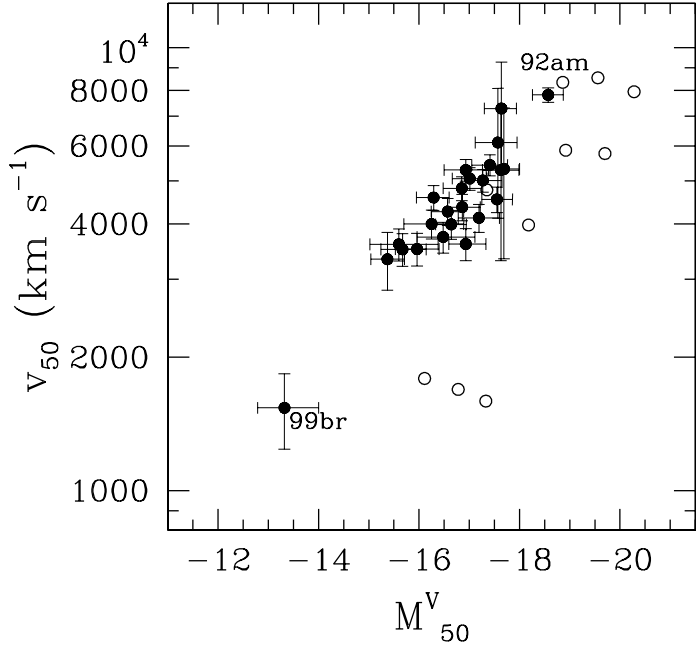}
\end{center}
\caption{
Envelope velocity versus absolute plateau $V$ magnitude for
24 SNe~IIP, both measured in the middle of the plateau (day 50)
(filled circles). The expansion velocities were obtained from
the minimum of the Fe II $\lambda$5169 lines. The absolute magnitudes
were derived from redshift-based distances and observed magnitudes
corrected for dust extinction. Open circles correspond to 
explosion models computed by Litvinova \& Nadezhin (1983, 1985)
for stars with $M_{ZAMS}$ $\geq$ 8 $M_\odot$. 
}
\label{L_v.fig}
\end{figure}

Figure \ref{L_v.fig} shows that, despite this diversity, 
the SN plateau luminosities are well correlated with the expansion velocities. 
Also shown with open circles are the explosion
models of Litvinova \& Nadezhin (1983, 1985, hereafter LN83 and LN85) for stars with
$M_{ZAMS}$ $\geq$ 8 $M_\odot$. It is clear that the luminosity-velocity
relation is also present in the theoretical calculations. This comparison
suggests that one of the main parameters behind this diversity is
the explosion energy, which causes great variation in 
the kinetic and internal energies. A similar result was recently
found by Zampieri et al. (2003b).

\begin{figure}
\begin{center}
\leavevmode\epsfxsize=10cm \epsfbox{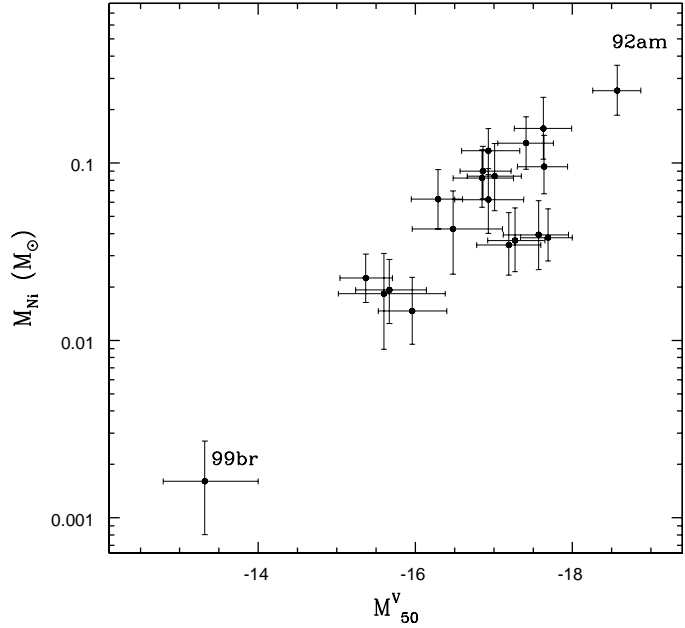}
\end{center}
\caption{Mass of $^{56}$Ni ejected versus
plateau luminosity measured 50 days after explosion.
}
\label{L_Ni.fig}
\end{figure}

In figure \ref{L_Ni.fig} I compare the luminosity during
the plateau and exponential phases. The latter is expressed in terms
of the mass of $^{56}$Ni ejected, $M_{Ni}$, assuming that the late-time lightcurve
is powered by the full trapping and thermalization of the $\gamma$ rays due to
$^{56}$Co $\rightarrow$ $^{56}$Fe ($^{56}$Co is the daughter of $^{56}$Ni,
which has a half life of only 6.1 days). There is clear evidence that SNe
with brighter plateaus also have brighter tails. A similar result
was recently found by Elmhamdi et al. (2003). Note that this correlation
is independent of the distance and reddening adopted for each SN.

The previous analysis shows that several regularities emerge among the
observed properties of SNe~IIP. Within the current uncertainties 
a single parameter is required to explain the variations in luminosity
and expansion velocity.

\section{Physical properties of Type II plateau supernovae}

Using hydrodynamic models, LN83 and LN85 derived approximate relations that
connect the explosion energy ($E$), the mass of the envelope ($M$), and the
progenitor radius ($R_0$) to three observable quantities, namely,
the duration of the plateau, the absolute $V$ magnitude, and the photospheric
velocity observed in the middle of the plateau. These formula
provide a simple and quick method to derive $E$, $M$, and $R_0$
from observations of SNe~II-P, without having to craft specific models for each SN.

\begin{figure}
\begin{center}
\leavevmode\epsfxsize=10cm \epsfbox{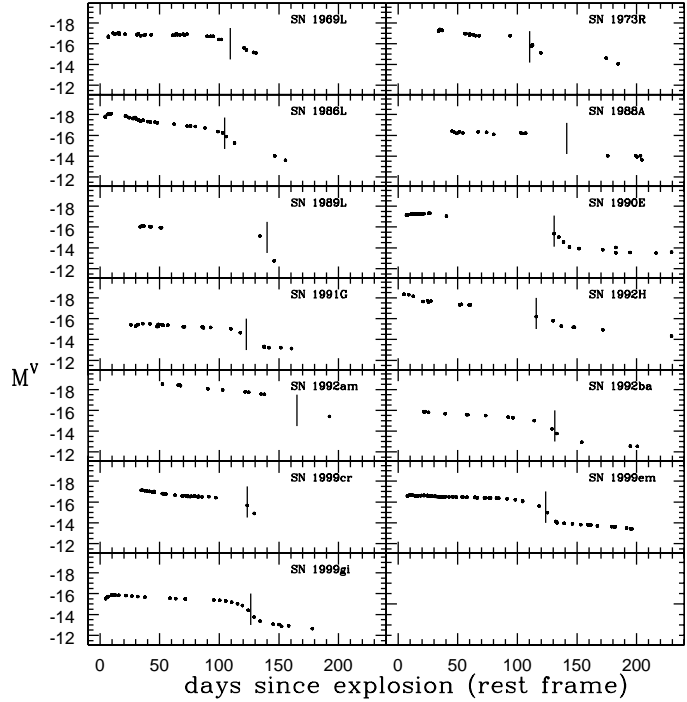}
\end{center}
\caption{Extinction corrected absolute $V$-band lightcurves of the 13 SNe~IIP.
The vertical bars indicate the end of the plateau phase for each supernova.
}
\label{MLC.fig}
\end{figure}

Of the 24 SNe~II-P considered above only 13 have sufficient data to apply
the method of LN85. The light curves for these SNe are shown in Fig. \ref{MLC.fig}.
The input parameters are listed in Table 3 of H03 and the output
parameters are summarized in Table \ref{SN.tab}. This table includes
physical parameters for 3 additional SNe available in the literature,
namely, SN~1987A (Arnett 1996), SN~1997D and SN~1999br (Zampieri et al. 2003a).
Although SN~1987A showed an atypical lightcurve due to the compact nature
of its blue supergiant progenitor, it was not fundamentally different than
ordinary SNe~II-P in the sense that it had a hydrogen-rich envelope at the
time of explosion. For this reason I include it in this analysis. To my
knowledge these are the only 16 SNe~IIP with available physical parameters.

\begin{table}[ht] 
\centering
\caption{Physical Parameters for Type II Supernovae.}
\begin{tabular}{lcccc}
\hline 
SN & Energy                & Ejected Mass & Initial Radius& References \\
   & (10$^{51}$ ergs)      & ($M_\odot$)  & ($R_\odot$)   &            \\
\hline
1969L  &  2.3$_{\rm -0.6}^{+0.7}$  & 28$_{\rm  -8}^{+11}$ & 204$_{\rm  -88}^{+150}$ & 1 \\
1973R  &  2.7$_{\rm -0.9}^{+1.2}$  & 31$_{\rm -12}^{+16}$ & 197$_{\rm  -78}^{+128}$ & 1 \\
1986L  &  1.3$_{\rm -0.3}^{+0.5}$  & 17$_{\rm  -5}^{ +7}$ & 417$_{\rm -193}^{+304}$ & 1 \\
1987A  &  1.7                      & 15                   & 42.8                    & 2 \\
1988A  &  2.2$_{\rm -1.2}^{+1.7}$  & 50$_{\rm -30}^{+46}$ & 138$_{\rm  -42}^{ +80}$ & 1 \\
1989L  &  1.2$_{\rm -0.5}^{+0.6}$  & 41$_{\rm -15}^{+22}$ & 136$_{\rm  -65}^{+118}$ & 1 \\
1990E  &  3.4$_{\rm -1.0}^{+1.3}$  & 48$_{\rm -15}^{+22}$ & 162$_{\rm  -78}^{+148}$ & 1 \\
1991G  &  1.3$_{\rm -0.6}^{+0.9}$  & 41$_{\rm -16}^{+19}$ &  70$_{\rm  -31}^{ +73}$ & 1 \\
1992H  &  3.1$_{\rm -1.0}^{+1.3}$  & 32$_{\rm -11}^{+16}$ & 261$_{\rm -103}^{+177}$ & 1 \\
1992am &  5.5$_{\rm -2.1}^{+3.0}$  & 56$_{\rm -24}^{+40}$ & 586$_{\rm -212}^{+341}$ & 1 \\
1992ba &  1.3$_{\rm -0.4}^{+0.5}$  & 42$_{\rm -13}^{+17}$ &  96$_{\rm  -45}^{+100}$ & 1 \\
1997D  &  0.9                      & 17                   & 128.6                   & 3 \\
1999br &  0.6                      & 14                   & 114.3                   & 3 \\
1999cr &  1.9$_{\rm -0.6}^{+0.8}$  & 32$_{\rm -12}^{+14}$ & 224$_{\rm  -81}^{+136}$ & 1 \\
1999em &  1.2$_{\rm -0.3}^{+0.6}$  & 27$_{\rm  -8}^{+14}$ & 249$_{\rm -150}^{+243}$ & 1 \\
1999gi &  1.5$_{\rm -0.5}^{+0.7}$  & 43$_{\rm -14}^{+24}$ &  81$_{\rm  -51}^{+110}$ & 1 \\
\hline
\end{tabular}
\label{SN.tab}
\\
Code: (1) Hamuy (2003); (2) Arnett (1996); (3) Zampieri et al. (2003a)\\
\end{table}

Among this sample, 9 SNe have explosion energies close to the canonical 1 foe value
(1 foe=10$^{51}$ ergs), 6 objects exceed 2 foes, and one has only 0.6 foes.
SN~1992am and SN~1999br show the highest and lowest energies with 5.5
and 0.6 foes, respectively. This reveals that SNe~II encompass a wide range in explosion energies.
The ejected masses vary between 14 and 56 $M_\odot$. Although the uncertainties are large it is interesting
to note that, while stars born with more than 8 $M_\odot$ can in principle undergo
core collapse, they do not show up as SNe~II-P. Perhaps they undergo significant
mass loss before explosion and are observed as SNe~IIn or SNe~Ib/c. It proves
interesting also that stars as massive as 50 $M_\odot$ seem able to retain a significant fraction
of their H envelope and explode as SNe~II. Objects with $M$$>$35$~M_\odot$ are
supposed to lose their H envelope due to strong winds, and become Wolf-Rayet
stars before exploding (Woosley et al. 1993). This result suggests that
stellar winds in massive stars are not so strong as previously thought, perhaps due to
smaller metallicities. Except for four objects, the initial radii vary between
114 and 586 $R_\odot$. Within the error bars these values correspond to those measured
for K and M red supergiants (van Belle et al. 1999),
which lends support to the generally accepted view that the progenitors of SNe~II-P have extended atmospheres at
the time of explosion (Arnett 1996). Three of the SNe~II-P of this sample, however, have
$R_0$$\sim$80 $R_\odot$ which corresponds to that of G supergiants.
This is somewhat odd because in theory such objects cannot have
plateau lightcurves but, instead, one like that of SN~1987A.
Note, however, that the uncertainties are quite large and it is possible that
these objects did explode as red supergiants.

\begin{figure}
\begin{center}
\leavevmode\epsfxsize=10cm \epsfbox{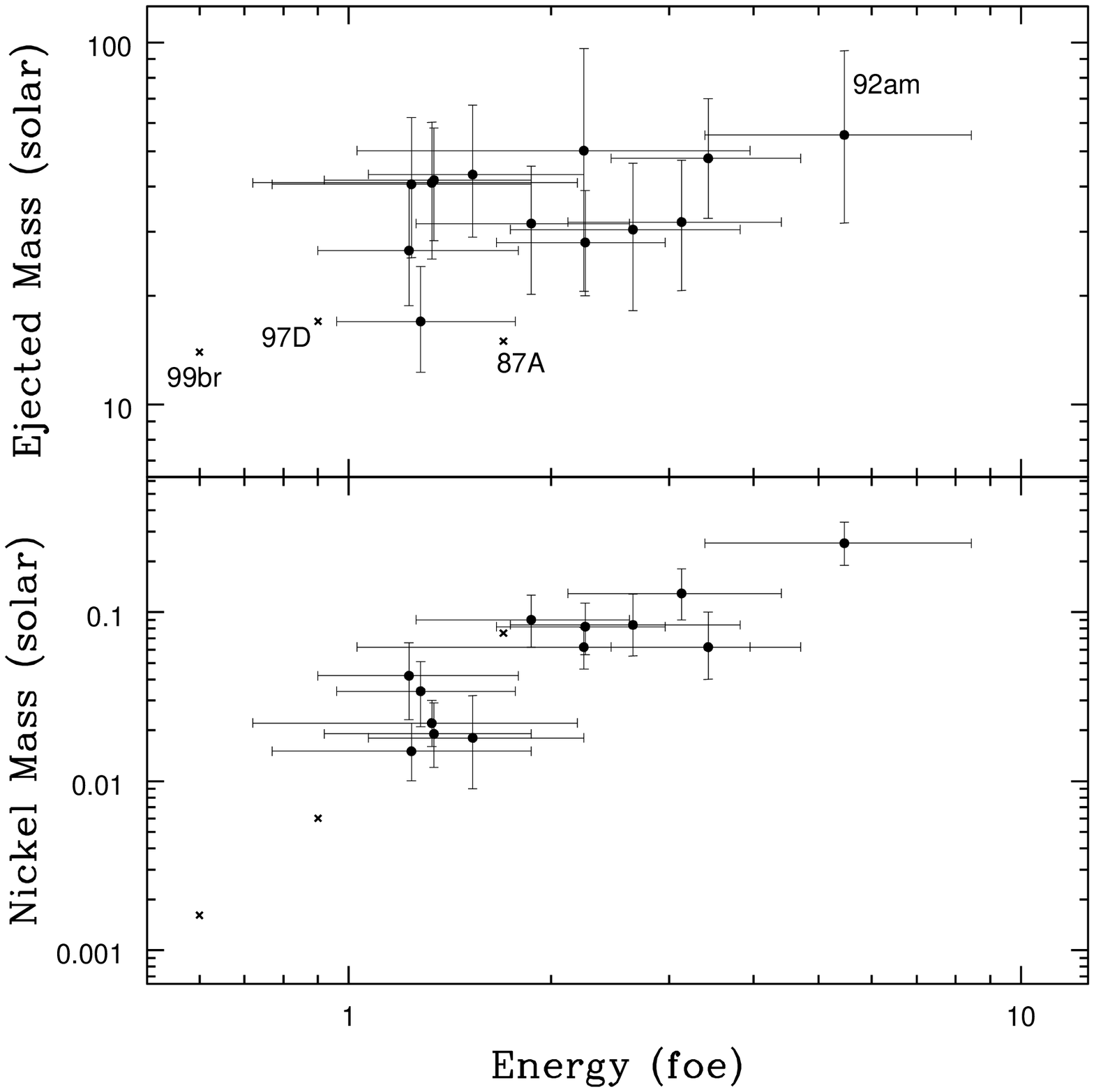}
\end{center}
\caption{
Envelope mass and nickel mass of SNe~II, as a function of explosion
energy. Solid points represent the 13 SNe~II-P for which I was able to apply
the technique of LN85. The three crosses correspond to SN~1987A,
SN~1997D, and SN~1999br which have been modeled in detail by Arnett (1996)
and Zampieri et al. (2003a). The nickel yield for SN~1999br comes from H03.
}
\label{ME.fig}
\end{figure}

Figure \ref{ME.fig} shows $M$ and $M_{Ni}$ as a function of $E$ for the 16 SNe~II-P.
Despite the large error bars, this figure reveals that a couple of correlations
emerge from this analysis. The first interesting result (top panel) is that the
explosion energy appears to be correlated with the envelope mass, in the sense
that more massive progenitors produce more energetic SNe. This suggests that stellar
mass plays a central role in the physics of core collapse. The second
remarkable result (bottom panel) is that SNe with greater energies produce more nickel
(a result previously suggested by Blanton et al. 1995).
This could mean that greater temperatures and more nuclear burning are reached in such SNe,
and/or that less mass falls back onto the neutron star/black hole in more energetic explosions.

\section{Conclusions}

\noindent 1) SNe~II-P encompass a wide range of $\sim$5 mag in plateau luminosities,
a five-fold range in expansion velocities, and a 100-fold range in tail luminosities.

\noindent 2) Despite this great diversity, SNe~II-P show several regularities such
as correlations between plateau luminosities, expansion velocities, and tail luminosities,
which suggests a one parameter family, at least to first order.

\noindent 3) There is a continuum in the properties of SNe~II-P from faint,
low-velocity, nickel-poor events such as SN~1997D and SN~1999br, and
bright, high-velocity, nickel-rich objects like SN~1992am. 

\noindent 4) SNe~IIP encompass a wide range in explosion energies (0.6-5.5 foes),
ejected masses (14-56 $M_\odot$), initial radii (80-600 $R_\odot$), and $^{56}$Ni
yields (0.002-0.3 $M_\odot$).

\noindent 5) Despite the large error bars, a couple of correlations emerge
from the previous analysis: (1) more $^{56}$Ni is ejected in SNe with greater
energies; (2) progenitors with greater masses produce more energetic explosions.
This suggests that the physics of the core collapse and the fate of massive stars
is, to a large extent, determined by the mass of the progenitor.

\bigskip\noindent
{\it Acknowledgements} Support for this work was provided by NASA through Hubble Fellowship
grant HST-HF-01139.01-A awarded by the Space Telescope Science Institute,
which is operated by the Association of Universities for Research in Astronomy,
Inc., for NASA, under contract NAS 5-26555.

\eject

\end{document}